\documentclass[aps,pre,reprint,superscriptaddress,amsmath,amssymb,showpacs,
floatfix]{revtex4-1}
\usepackage{graphicx}
\usepackage{bm}
\usepackage{color}

\newcommand{\be}{\begin{equation}}
\newcommand{\ee}{\end{equation}}
\newcommand{\bea}{\begin{eqnarray}}
\newcommand{\eea}{\end{eqnarray}}
\newcommand{\beax}{\begin{eqnarray*}}
\newcommand{\eeax}{\end{eqnarray*}}

\newcommand{\bmf}{{\bm{f}}}
\newcommand{\bmq}{{\bm{q}}}
\newcommand{\bmv}{{\bm{v}}}
\newcommand{\bmx}{{\bm{x}}}
\newcommand{\bmxi}{{\bm{\xi}}}
\newcommand{\bmdel}{{\bm{\nabla}}}
\newcommand{\bmD}{{\bm{D}}}
\newcommand{\bmB}{{\bm{B}}}

\newcommand{\erf}{\mathrm{erf}}

\newcommand{\R}{\mathbb{R}}

\newcommand{\sfF}{{\sf F}}
\newcommand{\sfQ}{{\sf Q}}

\newcommand{\apart}{streaming }
\newcommand{\spart}{down-hill }

\begin{document}

\title{On the steady state probability distribution 
of nonequilibrium stochastic systems}
\author{Jae Dong Noh}
\affiliation{Department of Physics, University of Seoul, Seoul 130-743,
 Korea}
\affiliation{School of Physics, Korea Institute for Advanced Study,
Seoul 130-722,  Korea}
\author{Joongul Lee}
\affiliation{Department of Mathematics Education, Hongik University, Seoul
121-791, Korea}

\date{\today}

\begin{abstract}
A driven stochastic system in a constant temperature heat bath 
relaxes into a steady state which is characterized by the steady state 
probability distribution.
We investigate the relationship between the driving force and the steady 
state probability distribution. We adopt the force decomposition method
in which the force is decomposed as the sum of a gradient of a steady 
state potential and the remaining part. The decomposition method allows one
to find a set of force fields each of which is compatible to a given
steady state. Such a knowledge provides a useful insight on stochastic
systems especially in the nonequilibrium situation. 
We demonstrate the decomposition method in stochastic systems 
under overdamped and underdamped dynamics and discuss the connection between
them.
\end{abstract}
\pacs{05.10.Gg, 05.70.Ln, 05.40.Jc, 05.40.-a}
% 05.70.Ln : Nonequilibrium irreversible thermodynamics
% 05.10.Gg : Stochastic analysis methods
% 05.40.-a : Fluctuation phenomena, random processes, noise, and Brownian
%            motion
% 05.40.Jc : Brownian motion
\maketitle

\section{Introduction}\label{sec:intro}
Systems in contact with a thermal heat bath are described by the 
probability distribution~\cite{Risken:1996vl,Gardiner:2010tp,VanKampen:2011vs}. 
When the dynamics satisfies the detailed balance~(DB), a system evolves into 
the equilibrium state whose probability distribution is
given by the Boltzmann distribution. Recently, much attention has been paid
to nonequilibrium systems since many interesting small-sized systems, 
biological systems, and complex systems are usually driven out of
equilibrium. In contrast to equilibrium systems, the DB is broken in 
nonequilibrium systems. Hence the steady state distribution deviates from the 
Boltzmann distribution. It is one of the important tasks of nonequilibrium
statistical mechanics to characterize the nonequilibrium steady state.

The exact steady state probability distribution is known for only a few cases.
Some of stochastic systems governed by the master equation, such as the
asymmetric simple exclusion process~\cite{Derrida:1998vb} and the zero range
process~\cite{SPITZER:1970vp,Evans:2000vw,Noh:2005io}, 
are exactly solvable. For overdamped Langevin equation systems, 
it is easily found on a one-dimensional ring~\cite{Risken:1996vl}. 
In higher dimensions with a linear force, 
the probability distribution can be written in terms of an anti-symmetric
matrix whose elements are given by the solution of a set of algebraic
equations~\cite{Kwon:2005uq}. For underdamped Langevin equation systems, 
several classes of solvable models are found in one spatial
dimension~\cite{Wang:1998wp,Wang:2000ug}. Besides the exceptional solvable
cases, it is hard to obtain explicitly the probability distribution of
general nonequilibrium systems. 

In this paper, we consider 
stochastic systems which are in thermal contact with a single heat 
bath and driven by an external force. 
Given the difficulty of finding the steady state, we focus
on the algebraic relationship between the driving force and the steady state
probability distribution. Our approach is based on the decomposition of the
driving force into two parts. The decomposition has been recognized 
as a useful tool to characterize the steady 
state~\cite{{Risken:1996vl},{Qian:1998uz},{Kwon:2005uq},{Kwon:2011ct}}.
Once the steady state is known, the force can be uniquely decomposed into two,
each of which reflects the probability current in the steady state.

Applicability of the decomposition method has been limited because 
it requires the knowledge of the steady state probability
distribution in advance. 
We elaborate more on the decomposition method to reveal some lights it can
shed on the steady state of general stochastic systems. Our findings are
listed as below:
For the overdamped dynamics, it yields a first-order nonlinear differential 
equation for the steady state distribution in comparison with the  
second-order linear differential equation obtained from 
the Fokker-Planck equation.
It allows us to find a solvability condition for a linear diffusion system. 
When the force matrix is {\em normal}, the steady state probability 
distribution is found explicitly.
For general stochastic systems, overdamped or underdamped, 
it also provides a systematic way 
to find a class of driving forces that share the same steady state. Such an
information is useful in understanding the extent to which a given steady
state covers different physical systems as well as the allowed forms of 
the steady state probability distribution. This study shows how 
overdamped dynamics is achieved from underdamped dynamics.
We also find interesting solvable nonequilibrium 
models which can be used to examine the recent developments such as the
modified fluctuation dissipation
relations~\cite{{Seifert:2010ba},{Prost:2009ky},{Verley:2011kb}} 
and the nonequilibrium fluctuation
theorems~\cite{{Jarzynski:1997uj},{Crooks:1999ta},{Noh:2012hg},{Lee:2013fb},{Spinney:2012di}}.

The paper is organized as follows. In Sec.~\ref{sec:fd}, we review 
the Langevin equation and the Fokker-Planck equation formalisms and then 
introduce the force decomposition method. It is then applied to the
overdamped systems in Sec.~\ref{sec:over} and to the underdamped systems in
Sec.~\ref{sec:under}. We conclude the paper with summary in
Sec.~\ref{sec:summ}.

%%%%%%%%%%%%%%%%% section II %%%%%%%%%%%%%%%%
\section{Force decomposition}\label{sec:fd}
In this section, we briefly review the Langevin equation and the
Fokker-Planck equation formalisms in order to set the notation. 
We refer the reader to 
Refs.~\cite{Risken:1996vl,Gardiner:2010tp,VanKampen:2011vs} 
for the detailed review.
The Langevin equation for a stochastic system with $n$ real variables
$\bmq=(q_1,q_2,\cdots,q_n)^T\in \R^n$, represented as a column vector, 
is written as
\be\label{Langevin_gen}
\dot{q}_i = h_i(\bmq) +  \sum_j g_{ij} \zeta_j(t) \ ,
\ee
where $\bm{\zeta}(t)=(\zeta_1(t),\cdots,\zeta_n(t))^T$ are the Gaussian 
random noises satisfying
\be
\langle \zeta_i(t) \rangle = 0, \ \langle \zeta_i(t)\zeta_j(t') \rangle = 2
\delta_{ij} \delta(t-t') \ ,
\ee
$h_i$'s represent the deterministic part of the time 
evolution, and $g_{ij}$'s represent the noise strength. 
The superscript ${}^T$ denotes the transpose.
The variable $q_i$ may represent a component of the position or the velocity
vector. 
We are interested in the steady state of a stochastic system 
in a thermal heat bath, so both $h_i$ and $g_{ij}$ are assumed to be
independent of  $t$ and $g_{ij}$ does not depend on $\bmq$.

Let $P(\bmq,t)$ denote the probability distribution of the system at time $t$.
It evolves in time following 
the Fokker-Planck equation~\cite{Risken:1996vl}
\be\label{FP_gen}
\frac{\partial}{\partial t}P(\bmq,t) =
\left[ - \sum_{i=1}^{n} \frac{\partial}{\partial q_i} D_i
+ \sum_{i,j=1}^{n} \frac{\partial^2}{\partial q_i
\partial q_j} D_{ij} \right] P ,
\ee
where $\bm{D}=(D_1,\cdots,D_{n})^T$ is the drift vector with
\be
D_i = h_i(\bmq)
\ee
and $\mathsf{D} = \{D_{ij}\}$ is the diffusion matrix with
\be
D_{ij} = \sum_k g_{ik} g_{jk} \ .
\ee
The Fokker-Planck equation can be casted into the continuity equation 
\be\label{continuity_eq}
\frac{\partial P}{\partial t} + \sum_{i=1}^n \frac{\partial J_i}{\partial
q_i} = 0
\ee
with the probability current density
\be\label{current_density}
J_i(\bmq,t) = \left( D_i - \sum_j \frac{\partial}{\partial q_j} D_{ij} \right)
P(\bmq,t) \ .
\ee

When $t$ goes infinity, the system reaches the steady state. The steady
state probability distribution will be denoted as
\be\label{Pst_def} P_{st}(\bmq) = e^{-\phi(\bmq)} \ee  
with the steady state potential $\phi(\bmq)$. 
Obviously, it is given by the solution of
\be\label{steady_state_cond}
\left[ - \sum_{i=1}^{n} \frac{\partial}{\partial q_i} D_i
+ \sum_{i,j=1}^{n} \frac{\partial^2}{\partial q_i
\partial q_j} D_{ij} \right] P_{eq}(\bmq) = 0 \ .
\ee
The steady-state probability current density is given by
\be\label{Jst}
J_{st,i}(\bmq) = e^{-\phi(\bmq)}
\left(D_i + \sum_j D_{ij}
\frac{\partial \phi}{\partial q_j} \right) \ ,
\ee
where we used that the $D_{ij}$'s are independent of $\bmq$.
The steady-state condition of \eqref{steady_state_cond} becomes 
\be\label{d_free} 
\sum_i \frac{\partial J_{st,i}}{\partial q_i} =0.
\ee

The expression \eqref{Jst} suggests that the drift coefficient 
may be decomposed as $D_i = D_i^{(s)} + D_i^{(a)}$ with~\cite{Risken:1996vl}
\bea
D_i^{(s)}(\bmq) &=& 
        - \sum_{j} D_{ij} \frac{\partial \phi}{\partial q_j} \ , \label{Ds} \\
D_i^{(a)}(\bmq) &=& D_i - D_i^{(s)} \ . \label{Da}
\eea
The steady-state current is determined by $\bmD_i^{(a)}$;  
\be\label{Jst_D}
J_{st,i}(\bmq) = D_i^{(a)}(\bmq) e^{-\phi(\bmq)} \ .
\ee
We will call $\bmD^{(a)}=(D_1^{(a)},\cdots,D_n^{(a)})^T$ and
$\bmD^{(s)}=(D_1^{(s)},\cdots,D_n^{(s)})^T$ the {\em \apart} vector 
and the {\em \spart} vector, respectively. 

As stated in Ref.~\cite{Risken:1996vl}, the decomposition is possible only 
when the steady state potential $\phi$ is known, which is hard in general. 
Despite the difficulty, we find that the decomposition is useful in studying 
the nature of the steady states. Using \eqref{d_free} and \eqref{Jst_D}, we
obtain the relation 
\be\label{decomp2}
\sum_i \left( \frac{\partial D_i^{(a)}}{\partial q_i} - D_i^{(a)}
\frac{\partial \phi}{\partial q_i} \right) = 0 \ .
\ee
that constrains the \apart part and the steady state potential.
We can approach the steady state problem in a different
perspective through the relation. 
These will be pursued for systems under the overdamped and the underdamped 
dynamics for a Brownian particle
coupled to a single heat bath in the following sections.

\section{Overdamped dynamics}\label{sec:over}
A Brownian particle in the $d$-dimensional space 
is driven by an external force in a thermal heat bath characterized by
the damping coefficient $\gamma$ and the temperature $T$. 
The overdamped dynamics is governed by the Langevin equation
\be\label{Langevin_eq_od}
\gamma \dot\bmx = \bmf(\bmx) + \bmxi(t)
\ee
where $\bmx = (x_1,\cdots,x_d)^T \in \R^d$ denotes the position vector
and $\bmf(\bmx)=(f_1(\bmx),\cdots,f_n(\bmx))^T$ denotes an external force. 
The thermal noise $\bmxi(t)=(\xi_1(t),\cdots,\xi_d(t))^T$ satisfies
\be
\langle \xi_i(t) \rangle = 0 , \ \langle \xi_i(t) \xi_j(t') \rangle =
2\gamma  k_B T \delta_{ij}\delta(t-t') \ .
\ee
This system corresponds to \eqref{Langevin_gen} with 
$n=d$, $\bmq=\bmx$, $h_i = f_i/\gamma$, and $g_{ij} =
\sqrt{k_B T/\gamma}\delta_{ij}$. 
In terms of the Fokker-Planck equation,
the drift coefficient and the diffusion matrix are given by
\be
D_i = \frac{f_i}{\gamma} \mbox{ and } 
D_{ij} = \frac{k_B T}{\gamma}\delta_{ij} \ .
\ee
Hereafter, the Boltzmann constant $k_B$ will be set to unity.

The drift vector is proportional to the driving force. 
Thus one can decompose the total force $\bmf(\bmx)$, instead of the drift
vector, as
\be\label{decomp3}
\bmf(\bmx) = \bmf_c(\bmx) + \bmf_{nc}(\bmx) \ ,
\ee
where 
\be\label{f_c}
\bmf_c(\bmx) = - T \bmdel \phi(\bmx) 
\ee
and
\be\label{f_nc}
\bmf_{nc}(\bmx) = \bmf(\bmx) + T \bmdel \phi(\bmx)  
\ee
with the gradient operator $\bmdel$.
Note that $\bmf_c = \gamma \bmD^{(s)}$ and $\bmf_{nc} = \gamma \bmD^{(a)}$.
From \eqref{d_free}, we require the steady state condition
\be\label{eq:dc}
\bmdel \cdot \left( e^{-\phi} \bmf_{nc}\right) = 0 \ .
\ee
It can be rewritten as 
\be\label{fd_od_1}
(\bmf_c + T \bmdel) \cdot \bmf_{nc} = 0 
\ee
or
\be\label{fd_od_2}
(\bmf - \bmf_{nc} + T \bmdel) \cdot \bmf_{nc} = 0 \ .
\ee
It is a nonlinear first-order partial-differential equation for 
$\bmf_{nc}(\bmx)$. Once the solution is found to a given force $\bmf(\bmx)$, 
$\bmf_c(\bmx) = \bmf(\bmx)-\bmf_{nc}(\bmx) = -T \bmdel \phi(\bmx)$ yields
the steady state probability distribution $P_{st}(\bmx)=e^{-\phi(\bmx)}$.

The decomposition is trivial when the force is conservative in the form of
$\bmf = -\bmdel V(\bmx)$ with a scalar function $V(\bmx)$. 
The choice of $\bmf_c = \bmf$ and $\bmf_{nc} =0$ satisfies 
\eqref{fd_od_1}. Hence,  
the steady state distribution is given by the
equilibrium Boltzmann distribution with 
$\phi(\bmx) = \beta V(\bmx)$ up to a normalization constant with $\beta=1/T$.
When the force is nonconservative, the decomposition becomes nontrivial.

Decomposition of a given vector field $\bmf(\bmx)$ is an interesting
mathematical problem. For instance, the Hodge
theory~\cite{griffiths2011principles,warner1971foundations}
guarantees that a doubly-periodic
vector field $\bmf(x,y)$ on the $xy$-plane, namely a vector field on a torus,
can be written uniquely as 
the sum of the three vector fields;
\be\label{Hodge}
\bmf(x,y) = \bmdel \psi(x,y) + \mathbf{k}\times \bmdel 
\varphi(x,y) +(c_1,c_2) \ ,
\ee
where $\psi$ and $\varphi$ are smooth doubly periodic functions on 
$\R^2$, $\mathbf{k}$ is the unit vector in the $z$ direction perpendicular 
to the $xy$-plane, $\times$ denotes the cross product, and $(c_1,c_2)$ is 
a constant vector field.
Note that the second and third types of vector fields are divergence free.
Let us consider the case when $\psi(\bmx)=c\varphi(\bmx)$ for a constant $c$, and
$c_1=c_2=0$. In this case, one can set
$\bmf_{c}=\bmdel \psi$ and $\bmf_{nc}= \mathbf{k}\times \bmdel \varphi$,
as they are pointwise perpendicular~($\bmf_c \cdot \bmf_{nc}=0$) 
and $\bmf_{nc}$ is divergence free~($\bmdel \cdot \bmf_{nc}=0$).
Such a system has the steady state potential $\phi(x,y) = -\beta \psi(x,y)$.

\subsection{Driven particle in one-dimensional ring}
The decomposition condition in \eqref{fd_od_2} can be solved exactly 
in one dimension.
Consider a particle in a one-dimensional ring $0\leq x\leq L$ which is 
subject to a periodic potential $V(x)=V(x+L)$ and a uniform driving force 
$f_0$ so that $f(x) = -V'(x) + f_0$. The prime $'$ denotes the derivative
with respective to $x$.
The periodic boundary condition is imposed.
This problem was studied thoroughly in e.g. 
Refs.~\cite{Risken:1996vl,Mehl:2008dw}. 

In the conventional approach~\cite{Risken:1996vl}, the steady state 
is found from the probability conservation. From
\eqref{continuity_eq} and \eqref{current_density}, the steady state current
\be\label{J_1d}
J_{st} = \left[ f(x) P_{st}(x) - T P_{st}'(x) \right] /\gamma
\ee
should be a constant independent of $x$ in one dimension. 
Introducing a multi-valued pseudo-potential function
\be
\widetilde{V}(x) = V(x) - f_0 x
\ee 
and multiplying the both sides of \eqref{J_1d} with 
an integrating factor $e^{\beta \widetilde{V}(x)}$, one obtains
\be
\left[e^{\beta\widetilde{V}(x)}P_{st}(x)\right]' = -\beta\gamma J_{st}
e^{\beta\widetilde{V}(x)} \ .
\ee
It is integrated to yield the solution
\be\label{Pst_1d_sol}
P_{st}(x) = \beta\gamma J_{st} e^{-\beta\widetilde{V}(x)} \left[ c - 
             \int_0^x e^{\beta\widetilde{V}(y)} dy \right]
\ee
with an integration constant $c$ which is determined from the
periodic boundary condition $P_{st}(0) = P_{st}(L)$:
\be\label{eq_c}
c = \frac{e^{-\beta\widetilde{V}(L)}\int_0^L e^{\beta\widetilde{V}(y)}dy}
         {e^{-\beta\widetilde{V}(L)}-e^{-\beta\widetilde{V}(0)}}
  = \frac{\int_0^L e^{\beta(V(y) - f_0 y)} dy}{1-e^{-\beta f_0 L}}
\ee
The steady state current $J_{st}$ is determined from the normalization 
$\int P_{st}(x) dx = 1$, which yields
\be
J_{st} = \frac{T/\gamma}{\int_0^L dx e^{-\beta(V(x)-f_0 x)} 
        \left( c - \int_0^x dy e^{\beta (V(y)-f_0y)}\right)}. 
\ee

The same steady state distribution is reproduced from \eqref{fd_od_2}, 
which becomes 
\be
T f_{nc}' - f_{nc}^2 + f(x) f_{nc} = 0 \ .
\ee
The transformation $g(x) = {1}/{f_{nc}(x)}$ linearizes the differential
equation to the form 
\be
- T g' + f g = 1 \ .
\ee
Note that this equation is almost the same as that for $P_{eq}(x)$ in
\eqref{J_1d}. Following the same procedure, we obtain that
\be
g(x) = \beta e^{-\beta \widetilde{V}(x)} \left[ c - \int_0^x e^{\beta
\widetilde{V}(y)} dy \right]
\ee
where $c$ is given in \eqref{eq_c}. The steady state distribution $P_{st}(x)
= e^{-\phi(x)}$ is then determined from the relation 
$f_c(x) = -T \phi'(x) = f(x)-f_{nc}(x) = f(x) - 1/g(x)$. Using the solution
for $g(x)$, we find that
\be
\phi(x) = \beta \widetilde{V}(x) - \ln \left| c - \int_0^x
e^{\beta\widetilde{V}(y)}dy \right| + \phi_0
\ee
with a normalization constant $\phi_0$. This solution is identical to the
one in \eqref{Pst_1d_sol}. This example shows that the decomposition method
works equally well as the conventional method for one-dimensional systems.

\subsection{Linear diffusion systems}
Consider a linear diffusion system where the force is linear in $\bmx \in
\R^d$:
\be\label{linear_force}
\bmf(\bmx) = \sfF \bmx \ ,
\ee
where $\sfF$ is a $d\times d$ force matrix whose elements are constant. 
This system is also called the $d$-dimensional Ornstein-Uhlenbeck process. 
It attracts a lot of recent interests for the study of nonequilibrium
fluctuations~\cite{Kwon:2005uq,Noh:2013jm,Kwon:2011fk}.
Our task is to find the steady
state distribution function $P_{st}(\bmx) = e^{-\phi(\bmx)}$ using the
decomposition method. 

We write the conservative part of \eqref{linear_force} as 
$\bmf_c = \sfF_c \bmx$ with a symmetric matrix $\sfF_c = \sfF_c^T$
and the nonconservative part as $\bmf_{nc} = \sfF_{nc} \bmx$ with
$\sfF_{nc} = \sfF - \sfF_{c}$. 
Then, the steady state condition \eqref{fd_od_1} becomes
\be\label{f_cond}
\bmx^T \sfF_c \sfF_{nc} \bmx = -T\; {\rm Tr}\; {\sfF_{nc}} \ .
\ee
This is valid at all $\bmx$ only when
$\sfF_{nc}$ is traceless and ${\sf A} \equiv \sfF_c \sfF_{nc}$ 
is anti-symmetric:
\be\label{mat_rel}
{\rm Tr} \sfF_{nc} = 0 \mbox{ and } \sfF_c \sfF_{nc} = - \sfF_{nc}^T \sfF_c .
\ee
This condition implies that the nonconservative part
$\bmf_{nc}$ is divergence free~($\bmdel \cdot 
\bmf_{nc}={\rm Tr}\; \sfF_{nc}=0$) and perpendicular to the conservative 
part~($\bmf_c \cdot \bmf_{nc}=0$). 
Once the decomposition is found, the steady state
distribution function is given by $P_{st}(\bmx) = e^{-\phi(\bmx)}$ with
\be
\phi(\bmx) = -\frac{1}{2T}\ \bmx^T \sfF_c \bmx
\ee
up to a normalization constant.

The decomposition condition may be found by solving the set of algebraic
equations in \eqref{mat_rel} for the elements of $\sfF_c$
and $\sfF_{nc}=\sfF-\sfF_{c}$. 
Kwon {\it et al}~\cite{Kwon:2005uq} considered general linear diffusion 
systems with nondiagonal diffusion matrix. Their formalism yields that the
conservative part is written in the form of $\sfF_c = (1+{\sfQ}/T)^{-1} \sfF$ 
with an anti-symmetric matrix $\sf{Q}$~\cite{Kwon:2005uq}. 
The anti-symmetric matrices $\sf{Q}$ and ${\sf A} = \sfF_c \sfF_{nc}$ 
are related as $\sfQ = T \sfF_c^{-1} \mathsf{A} \sfF_c^{-1}$.

The decomposition is trivial for equilibrium systems where the force is
conservative with a symmetric $\sfF$~($\sfF_c = \sfF$ and $\sfF_{nc} = 0$).
Besides the equilibrium case, the decomposition method allows one to
find another solvable class for the force matrix.
Suppose that the force matrix $\sfF$ is {\em normal}~\cite{{horn2012matrix}}, 
which means that
\be\label{normal}
\sfF \sfF^T = \sfF^T \sfF \ .
\ee
A matrix is then decomposed as
\be\label{d_for_normal}
\sfF_c = \frac{1}{2}(\sfF + \sfF^T) \mbox{ and } \sfF_{nc} =
\frac{1}{2}(\sfF-\sfF^T) \ .
\ee
Note that $\sfF_c$ is symmetric and $\sfF_{nc}$ is traceless. Furthermore,
the normal condition in \eqref{normal} guarantees that 
\be
\sfF_c \sfF_{nc} = \frac{1}{4} (\sfF + \sfF^T)(\sfF-\sfF^T)
                 = \frac{1}{4} (\sfF^2 - (\sfF^T)^2) 
\ee
should be anti-symmetric. Consequently, \eqref{d_for_normal} is the proper 
decomposition leading to
\be
\phi(\bmx) = -\frac{1}{2T}\ \bmx^T \frac{(\sfF + \sfF^T)}{2} \bmx
\ee
up to a normalization constant.

The normal matrix includes orthogonal~($\sfF \sfF^T = \sf{I}$), 
symmetric~($\sfF=\sfF^T$), anti-symmetric~($\sfF = -\sfF^T$) matrices, and
others. For instance, 
\be\label{fex}
\sfF = \left( \begin{array}{ccc} -1 & -1 & 0 \\ 0 & -1 & -1 \\ -1 & 0 & -1
\end{array} \right) \ .
\ee
is an example of the normal matrix. 
The corresponding force $\bmf = \sfF \bmx$ attracts the Brownian
particle toward the origin with an additional rotational driving.
Due to the normality, the steady state potential is given by $\phi(\bmx) =
-\bmx^T (\sfF+\sfF^T) \bmx /(4T) = (x^2+y^2+z^2+xy+yz+zx)/(2T)$.

\subsection{Force fields family}
We have considered the special cases where the decomposition can lead to the
steady state explicitly, which is not possible for general cases.
In this subsection, we propose a different perspective in which the
decomposition plays an interesting role.
Instead of solving for the steady state potential to a given 
force field $\bmf(\bmx)$, we try to find a force field that 
leads to a given steady state $P_{st}(\bmx)=e^{-\phi(\bmx)}$. 
The steady state probability distribution can be measured 
experimentally~\cite{Mehl:2012fw,Blickle:2007fa,Speck:2007gv}. 
It would be interesting if one could reconstruct a force field from a 
measurement.

To a given steady state distribution $P_{st}(\bmx) = e^{-\phi(\bmx)}$, 
the nonconservative part $\bmf_{nc}(\bmx)$ should satisfy the steady state
condition~\eqref{eq:dc}.
The solution is not unique. Let $\bm{B}(\bmx)$ be any divergence-free vector
field. Then, \eqref{eq:dc} suggests that the nonconservative force should be
of the form
\be
\bmf_{nc}(\bmx) = \lambda e^{\phi(\bmx)} \bm{B}(\bmx)
\ee
with an arbitrary constant $\lambda$. Hence, any system with a force field
\be\label{force_od}
\bmf(\bmx) = -T \bmdel \phi(\bmx) + \lambda e^{\phi(\bmx)} \bm{B}(\bmx)
\ee
shares the same steady state distribution. 

There exist 
infinitely-many~(depending on the choice of $\bm{B}(\bmx)$) 
one-parameter~(represented by $\lambda$) families of the force fields 
to a given steady state potential $\phi(\bmx)$.
The parameter $\lambda$ represents the strength of the nonequilibrium
driving. 
In Ref.~\cite{Noh:2014co}, it was shown that the systems with $\lambda$ and
$-\lambda$ are dual to each other with respect to time reversal~(see also
Ref.~\cite{Sasa:2014bb}).

\begin{figure}
\includegraphics[width=0.32\columnwidth]{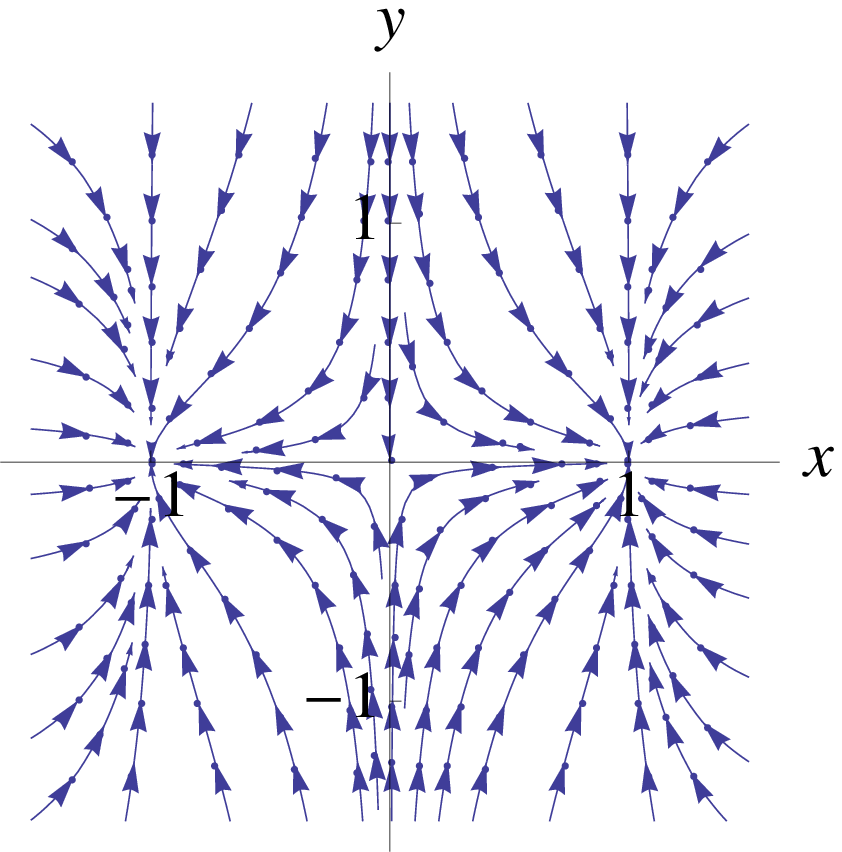}
\includegraphics[width=0.32\columnwidth]{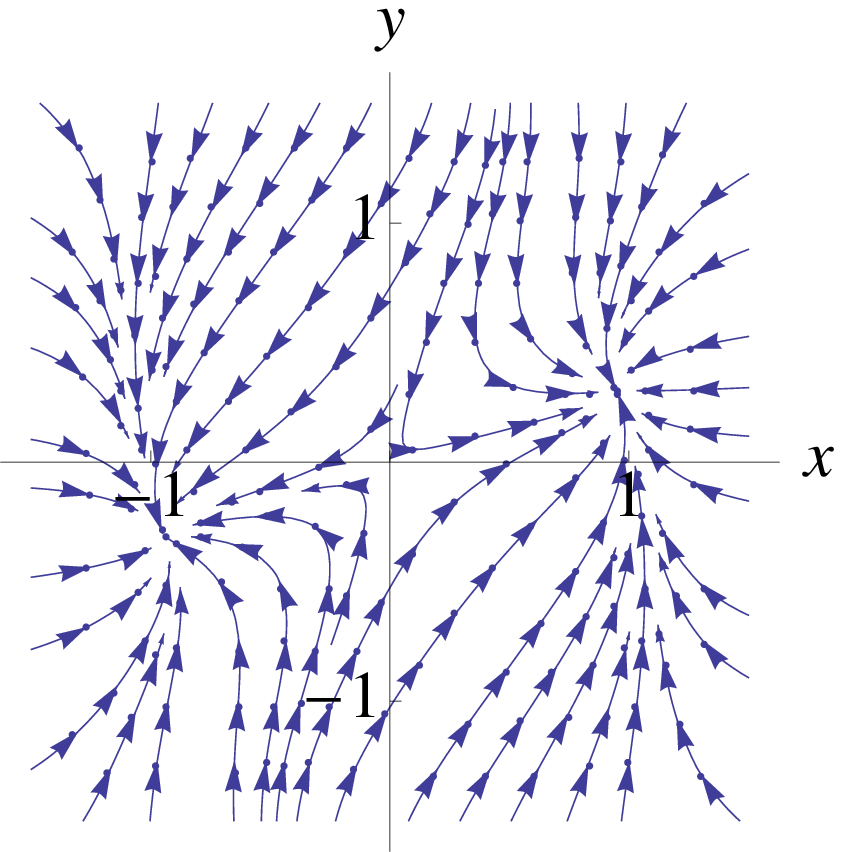}
\includegraphics[width=0.32\columnwidth]{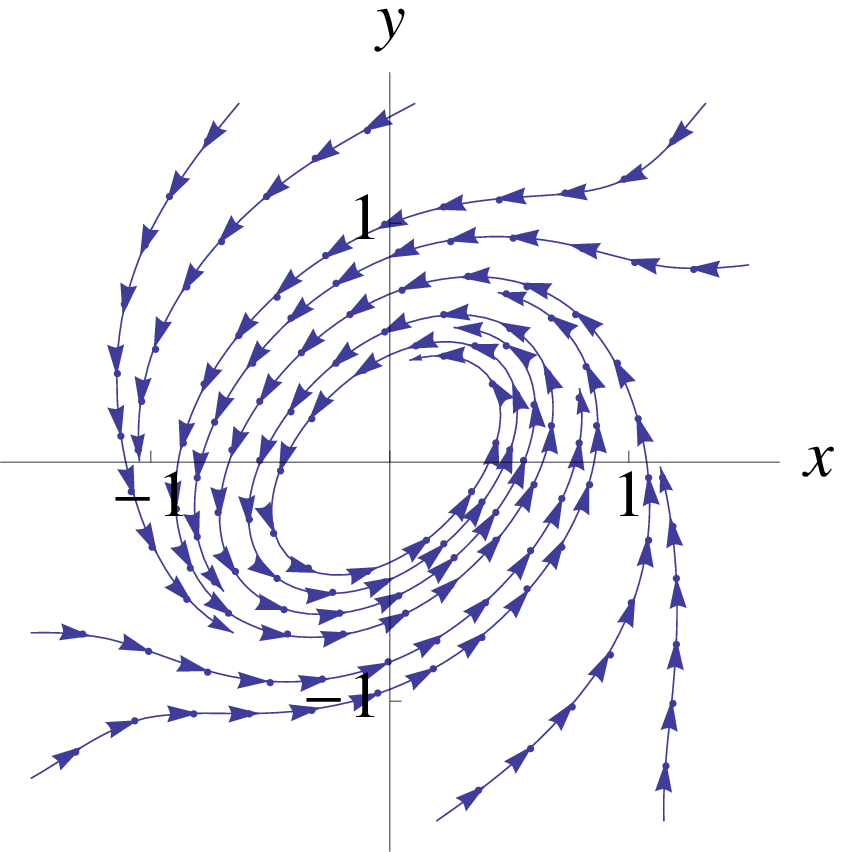}
\caption{Field lines of the force in \eqref{f_ex} with $\lambda=0.0$~(left), 
$0.5$~(center), and $2.0$~(right).}\label{fig2}
\end{figure}

\begin{figure}
\includegraphics*[width=0.8\columnwidth]{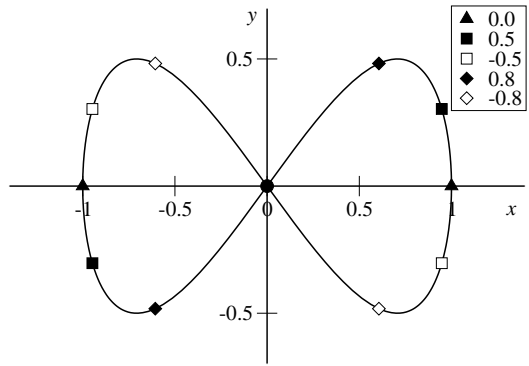}
\caption{Trajectory of the stable fixed points. The fixed points at several
values of $\lambda$ are marked with symbols. The origin $(0,0)$ marked by
the closed circle is the unique stable fixed point 
when $|\lambda|\geq 1$.}\label{fig3}
\end{figure}

In order to gain an intuitive understanding, 
consider a two-dimensional system having a steady state potential
\be\label{phi_ex}
\phi(x,y) = \beta \left[ \frac{1}{4}(x^2-1)^2 +\frac{y^2}{2}\right] +
\phi_0
\ee
with a normalization constant $\phi_0$. It is most probable to find the
particle at $(x^*,y^*)=(\pm 1,0)$.
Such a steady state is realized in an equilibrium system driven 
by a conservative force $\bmf(\bmx) = -T \bmdel \phi = (-x^3+x,-y)$.
The most probable positions coincide with the stable fixed point of the
conservative force.
As a divergence-free field, we choose 
\be
\bm{B}(\bmx) = e^{-\frac{r^4}{4}} 
   \left(\begin{array}{cc} -y \\ x \end{array} \right) 
\ee
with $r = \sqrt{x^2+y^2}$. 
Then, any total force of the form
\be\label{f_ex}
\bmf(\bmx) = \left( \begin{array}{cc} -x^3+x \\ -y \end{array} \right) +
             \lambda e^{-\frac{(r^2-1)^2+x^4+1}{4}} \left( \begin{array}{cc}
     -y \\ x \end{array} \right) 
\ee
has the same steady state potential in \eqref{phi_ex}.
The force field lines are drawn in Fig.~\ref{fig2} at a few values of
$\lambda$. One observes
a quantitative and qualitative changes as $\lambda$ varies. 
The stable fixed points move with $\lambda$ deviating from
the most probable points at $(x^*,y^*)$. The fixed points 
even undergo a 
bifurcation at $|\lambda|=1$, 
which leaves behind a single stable fixed point at $(x,y)=(0,0)$ 
for $|\lambda| > 1$~(see Fig.~\ref{fig3}). It is remarkable that those
force fields with different fixed point structures share 
the same steady state. The discrepancy between the most probable point and the
stable fixed point was reported with a perturbative calculation in
Ref.~\cite{Kwon:2011ct}. The decomposition method confirms that the
discrepancy is a general property of nonequilibrium systems driven by
a nonlinear force.

%%%%%%%%%%%%%%%%%%%%%%%% figures for the field lines %%%%%%%%%%%%%%%%%%%%%%%
%%%%%%%%%%%%%%%%%%%%%%%% figures for the field lines %%%%%%%%%%%%%%%%%%%%%%%

%%%%%%%%%%%%%%%%%%%%%%%% figures for the fixed points %%%%%%%%%%%%%%%%%%%%%%%
%%%%%%%%%%%%%%%%%%%%%%%% figures for the fixed points %%%%%%%%%%%%%%%%%%%%%%%

\section{Underdamped dynamics}\label{sec:under}
The underdamped dynamics of a Brownian particle of mass $m$ is governed by the
Langevin equation 
\bea
\dot{\bmx} &=& \bmv \\
m \dot{\bmv} &=& -\gamma \bmv + \bmf(\bmx,\bmv) + \bmxi
\eea
for the position $\bmx = (x_1,\cdots,x_d)^T\in \R^d$ and the velocity 
$\bmv=(v_1,\cdots,v_d)^T\in \R^d$. The force $\bmf$ may depend on both $\bmx$
and $\bmv$ in general. 
For convenience, we introduce a notation 
$\bm{q} = (q_1,\cdots,q_{2d})^T=(\bmx,\bmv)^T$ where $q_i = x_i$ 
and $q_{d+i} = v_i$ for $1\leq i\leq d$.
Then, the Langevin equations take the form of \eqref{Langevin_gen}. The
corresponding Fokker-Planck equation has the drift coefficients 
\be
D_i = D_{x_i} = v_i \mbox{ and } D_{d+i} = D_{v_i} = -\frac{\gamma}{m}v_i +
\frac{f_i}{m} 
\ee
with $1\leq i \leq d$. It is represented as a $(2d)$-dimensional column vector
\be
\bmD = \begin{pmatrix} \bmD_\bmx \\ \bmD_\bmv \end{pmatrix}
     = \begin{pmatrix} \bmv \\ -\frac{\gamma}{m}\bmv + \frac{1}{m}\bmf
\end{pmatrix} . 
\ee
The diffusion matrix $\mathsf{D} = \{D_{ij}\}$ has the 
elements $D_{ij} = D_{i,d+j} = D_{d+i,j} = 0$ and 
\be
D_{d+i,d+j} = D_{v_i, v_j} = \frac{\gamma  T}{m^2} \delta_{ij} 
\ee
for $1\leq i,j \leq d$. It is represented as a $(2d)\times(2d)$
matrix
\be
\mathsf{D} = \begin{pmatrix} \bm{0} & \bm{0} \\ \bm{0} & \frac{\gamma 
T}{m^2} \bm{1} \end{pmatrix} \ , 
\ee
where $\bm{0}$ and $\bm{1}$ denote the $d\times d$ null and identity
matrices, respectively.

We want to find a force field $\bmf(\bmx,\bmv)$ that
leads to a given steady state $P_{st}(\bmx,\bmv) = e^{-\phi(\bmx,\bmv)}$.
According to \eqref{Ds} and \eqref{Da}, we decompose the drift coefficient 
as the sum of
\bea
\bmD^{(s)} &=& - \mathsf{D} \bmdel \phi = \begin{pmatrix} \bm{0} \\ -\frac{\gamma  T}{m^2}
\bmdel_\bmv \phi \end{pmatrix} \label{Ds_ud} \\
\bmD^{(a)} &=& \begin{pmatrix} \bmD_\bmx^{(a)} \\
                               \bmD_\bmv^{(a)} \end{pmatrix}
            = \begin{pmatrix} \bmv \\ -\frac{\gamma}{m}\bmv+\frac{1}{m}\bmf
+\frac{\gamma T}{m^2} \bmdel_\bmv \phi\end{pmatrix},  \label{Da_ud}
\eea
where $\bmdel$ is the gradient operation in the $(\bmx,\bmv)$ space while
$\bmdel_\bmv$ is the gradient operator acting on the subspace of $\bmv$.
The steady state condition requires that
$$
\bmdel \cdot (\bmD^{(a)} e^{-\phi}) = \bmdel_\bmx \cdot
(\bmD_\bmx^{(a)}e^{-\phi}) + \bmdel_\bmv \cdot (\bmD_\bmv^{(a)} e^{-\phi})= 0 ,
$$
where $\bmdel_\bmx$ is the gradient operator acting on the subspace of $\bmx$.

The general solution for $\bmD^{(a)}$ is written as 
\be
\bmD^{(a)} = e^{\phi} \bmB
\ee
where $\bmB(\bmx,\bmv) \equiv \begin{pmatrix} \bmB_\bmx \\ \bmB_\bmv 
\end{pmatrix}$ is a divergence-free~($\bmdel_\bmx \cdot \bmB_\bmx 
+ \bmdel_\bmv \cdot \bmB_\bmv=0$) field. 

In contrast to the overdamped case, the auxiliary field $\bmB$ is subject to
the additional kinetic constraint that $\bmD^{(a)}_\bmx=\bmv$, which sets
$\bmB_\bmx = e^{-\phi} \bmv$. 
Hence, the divergence-free condition becomes
\be\label{Bv}
\bmdel_\bmv \cdot \bmB_\bmv = - \bmdel_\bmx \cdot \bmB_\bmx = \bmv \cdot
(\bmdel_\bmx \phi) e^{-\phi} \ .
\ee
With $\bmB_\bmv$ satisfying \eqref{Bv}, the force field is given by
\be\label{force_res}
\bmf(\bmx,\bmv) = \gamma \bmv - \frac{\gamma T}{m} \bmdel_\bmv \phi + m
\bmB_\bmv e^{\phi} \ .
\ee
In terms of $\bmD_\bmv^{(a)}=\bmB_\bmv e^{\phi}$, 
\eqref{Bv} and \eqref{force_res} become
\be\label{BvD}
\bmdel_\bmv \cdot \bmD_\bmv^{(a)} - \bmD_{\bmv}^{(a)} \cdot 
(\bmdel_\bmv \phi)  = \bmv \cdot (\bmdel_\bmx \phi)
\ee
and
\be\label{force_resD}
\bmf(\bmx,\bmv) = \gamma \bmv - \frac{\gamma T}{m} \bmdel_\bmv \phi + 
                  m \bmD_\bmv^{(a)} \ .
\ee

We compare the overdamped and the underdamped cases. In the former case, the
total force is decomposed into the conservative and nonconservative parts.
The nonconservative part $\bmf_{nc}$ is determined up to an arbitrary
parameter $\lambda$~(see \eqref{force_od}). Thus, any nonequilibrium system
characterized by a finite $\lambda$ finds the corresponding equilibrium 
system~($\lambda=0$) that shares the same steady state.
In the latter case, the decomposition does not separate the force into the
sum of conservative and nonconservative parts. Since \eqref{BvD} is
an inhomogeneous equation, $bmD_\bmv^{(a)}$ is given by the sum of the
homogeneous solution up to a multiplicative factor and the specific particular
solution. Hence, the correspondence between the steady state and
the driving force is more restrictive in the underdamped dynamics due to the
particular solution.
The correspondence is investigated further in the following subsections.

\subsection{Boltzmann distribution}
When the Brownian particle is driven by a velocity-independent 
conservative force $\bmf(\bmx) = - \bmdel_\bmx V(\bmx)$, the system reaches
the equilibrium Boltzmann distribution 
\be\label{Boltzmann_dist}
\phi(\bmx,\bmv) = \beta \left[\frac{m}{2} \bmv^2+V(\bmx)\right] \ .
\ee 
It is interesting to see whether the Boltzmann distribution
\eqref{Boltzmann_dist} can be also realized in other forces than the
conservative forces. 

For the Boltzmann distribution, $\bmdel_\bmv\phi = \beta m \bmv$ and 
$\bmD_\bmv^{(a)} = \bmf/m$ from \eqref{force_resD}. 
Hence, the force should satisfy
\be\label{f_BD}
\bmdel_\bmv \cdot \bmf - \beta m \bmv \cdot \bmf = m \beta \bmv \cdot 
(\bmdel_\bmx V(\bmx)) \ .
\ee
The general solution is given by
\be\label{f_BD_sol}
\bmf(\bmx,\bmv) = \lambda e^{\beta\frac{m \bmv^2}{2}} \bm{C}(\bmx,\bmv) 
- \bmdel_\bmx V(\bmx) \ ,
\ee
where $\bm{C}(\bmx,\bmv)\in \R^d$ is a vector field satisfying
$\bmdel_\bmv \cdot \bm{C}(\bmx,\bmv)$=0 and $\lambda$ is an arbitrary
parameter. 
The first term is the homogeneous solution and the second
term is the particular solution of \eqref{f_BD}.

One lesson from \eqref{f_BD_sol} is that there exist infinitely
many nonequilibrium forces sharing the same Boltzmann distribution
in the steady state. We also find that the Boltzmann distribution requires a
velocity dependent force except for the equilibrium case with $\lambda=0$.
If one perturbs an equilibrium system with any velocity-independent 
nonconservative force, the steady state must deviate from the Boltzmann
distribution. This is in sharp contrast to the overdamped system.

A special type of the homogeneous solution of \eqref{f_BD} is found by 
requiring $\bmdel_\bmv\cdot \bmf = \beta m \bmv \cdot \bmf=0$. It yields 
\be\label{f_BD_sol2}
\bmf(\bmx,\bmv) = \lambda \mathsf{H}(\bmx) \bmv - \bmdel_\bmx V(\bmx) \ ,
\ee
where $\mathsf{H}(\bmx)$ is an arbitrary antisymmetric matrix.
This force is linear in $\bmv$ and perpendicular to
$\bmv$. In $d=3$ dimensions, it is the familiar magnetic force on 
a charged particle in an inhomogeneous magnetic field.

\subsection{Shifted Boltzmann distribution in one-dimensional ring}
As a solvable extension, we consider 
a shifted Boltzmann distribution 
\be\label{model_phi}
\phi(x,v) = \beta \left[ \frac{m}{2} (v - \omega(x))^2 +
V(x) \right] 
\ee
in a one-dimensional ring of circumference $L$. The functions
$\omega(x+L)=\omega(x)$ and $V(x+L)=V(x)$ are periodic.
This form might be a natural extension of steady state potential from 
the overdamped dynamics to the underdamped dynamics.
In order to find a corresponding force field $f(x,v)$, 
one needs to solve~(see \eqref{Bv}) 
\be
\frac{\partial B_v}{\partial v} = \beta v \left[ -m (v-\omega)\omega' +
V'\right] e^{-\frac{\beta m}{2}(v-\omega)^2 -\beta V} \ .
\ee
Integrating the equation with respect to $v$, one obtains 
\bea
B_v &=& C(x) e^{-\beta V} + \left(\omega' v - \frac{V'}{m}\right) e^{-\phi}
\nonumber \\
    && + \sqrt{\frac{\pi}{2\beta m}}
 \left(\beta \omega V'-\omega'\right) \left[ 1 + \erf(Z)\right] e^{-\beta V} 
\eea
where $Z\equiv \sqrt{\frac{\beta m}{2}} (v-\omega(x))$,
$\erf(x) \equiv \frac{2}{\sqrt{\pi}}\int_0^x e^{-u^2}du$ is the error
function, and $C(x)$ is an arbitrary function of $x$ introduced as 
an integration constant. 
Hence, \eqref{force_res} yields 
\bea\label{force_field}
f(x,v) &=& m e^{Z^2} \left[ C(x) + \sqrt{\frac{\pi}{2\beta m}}
(\beta \omega V'-\omega') \left\{1+\erf(Z)\right\} \right] \nonumber \\
&& + (\gamma \omega + m v \omega' - V')  \ .
\eea

The factor $e^{Z^2}$ would drive the velocity to infinity. Such an
instability is avoided by taking $C(x)=0$ and 
\be\label{VandOmega}
\omega(x) V'(x) = T {w'(x)} \ ,
\ee
which leads to $V(x) = T \ln |\omega(x)| + V_0$
with an arbitrary constant $V_0$.
Therefore we conclude that the Brownian particle driven by the force
\be\label{solvable_force}
f(x,v) = \gamma w(x) - V'(x) + m v w'(x) 
\ee
with the constraint \eqref{VandOmega}
has the steady state potential given in \eqref{model_phi}.

When $\omega(x)=0$, the system corresponds to a free Brownian particle.
When $\omega(x)=\omega_0$ is a constant, $f(x,v) = \gamma \omega_0$ and
$\phi(x,v) = \beta m(v-\omega_0)^2/2$. This system corresponds to a Brownian
particle driven by a spatially uniform driving force on a ring. 
With position-dependent $\omega(x)$, the solvable force
should be velocity dependent.
The solvability requires that the $v$-dependent part 
and the $v$-independent part in \eqref{solvable_force} 
are interwoven intimately.
The solvable model could be useful in
testing various theoretical concepts such as the fluctuation theorems
for nonequilibrium systems, especially with velocity-dependent
forces~\cite{{Spinney:2012di},{Lee:2013fb}}.

\subsection{Overdamped limit}
The shifted Boltzmann distribution gives a hint how an underdamped system
is related to an overdamped system. Overdamped dynamics is achieved
by taking $m\to 0$ limit~\cite{Ao:2007ef,Durang:2013uf}. 
In this limit, the force \eqref{solvable_force} in the previous subsection 
becomes $f_{od}(x) = \gamma \omega(x) - V'(x)$, which is independent of $v$.
Comparing \eqref{VandOmega} with \eqref{fd_od_1}, we finds that $f_{nc}(x) =
\gamma \omega(x)$ and $f_{c}(x) = -V'(x)$ is the proper decomposition 
of the force $f_{od}$ in the overdamped limit. Hence, the overdamped system
with the force $f_{od}(x)$ has the steady state potential $\phi_{od}=\beta V$.
Interestingly, it is the same as the steady state potential in
\eqref{model_phi} after being averaged over $v$.

We can generalize the shifted Boltzmann distribution in arbitrary
$d$ dimensions. Consider a steady state potential of the form
\be\label{model_phi_d}
\phi(\bmx,\bmv) = \beta \left[ \frac{m}{2} (\bmv - \bm{\omega}(\bmx))^2 +
V(\bmx)\right]
\ee
with a vector field $\bm{\omega}(\bmx) = (\omega_1,\cdots,\omega_d)^T\in 
\R^d$ and a scalar field $V(\bmx)$. 
Repeating the similar algebra as in the previous subsection, 
we find that the auxiliary vector field 
$\bmB_\bmv(\bmx,\bmv)=(B_{v_1},\cdots,B_{v_d})$ is given by
\bea
B_{v_i} &=& C_i + 
\left[ (\bmv \cdot \bmdel_\bmx) \omega_i
- \frac{1}{m}\frac{\partial V}{\partial x_i}\right] e^{-\phi} \nonumber \\
&&+ \sqrt{\frac{\pi}{2\beta m}}
 \left(1+\erf{Z_i}\right) \Upsilon e^{-\phi+Z_i^2} \ ,
\eea
where 
\be\label{upsilon}
\Upsilon \equiv \left( \beta \bm{\omega}\cdot \bmdel_\bmx V - 
\bmdel_x\cdot \bm{\omega}\right) , 
\ee
$Z_i=\sqrt{\frac{\beta m}{2}}(v_i-\omega_i)$, and
$\bm{C}(\bmx,\bmv)=(C_1,\cdots,C_d)^T$ is any vector field satisfying
$\bmdel_\bmv \cdot \bm{C}=0$. It is straightforward to check this is 
the solution of \eqref{Bv}.
Thus, from \eqref{force_res}, the corresponding force is given by
\bea
f_i &=& \left[\gamma \bm{\omega}(\bmx) - \bmdel_\bmx V + m (\bmv \cdot
\bmdel_\bmx) \bm{\omega}  \right]_i \nonumber \\
&&+ C_i e^\phi + \sqrt\frac{\pi m}{2\beta} (1+\erf Z_i) \Upsilon  e^{Z_i^2}
\ .
\eea
In order to avoid instability, we will set $C_i=0$ and
$\Upsilon=0$. Consequently, the force is given by
\be\label{solvable_force_d}
\bmf = \gamma \bm{\omega}(\bmx) - \bmdel_\bmx V(\bmx) + m (\bmv \cdot
\bmdel_\bmx) \bm{\omega} 
\ee
with the constraint that $\Upsilon=0$.

Let us consider the overdamped limit~($m\to 0$). Then, the force becomes
$\bmf_{od} = \gamma \bm\omega - \bmdel_\bmx V$ which is independent of
$\bmv$. The stability condition $\Upsilon=0$
guarantees that $\bmf_c = -\bmdel_\bmx V$ and $\bmf_{nc} = \gamma \bm\omega$ 
is the proper decomposition of $\bmf_{od}$ satisfying \eqref{fd_od_1}. 
Hence, the steady state potential of the overdamped system is
given by $\phi_{od}(\bmx) = \beta V(\bmx)$. This is indeed the same steady 
state potential obtained from \eqref{model_phi_d}. 
It suggests that the shifted Boltzmann distribution is
a good approximation of the steady state potential of an underdamped system 
in the small $m$ limit.
The difference between the overdamped dynamics and the underdamped dynamics
lies in the $\bmv$-dependent term $(\bmv\cdot\bmdel_\bmx)\bm\omega$.
The origin of this force and its implications are left for a future study.

\section{Summary}\label{sec:summ}
We have investigated implications of the decomposition method on the 
relationship between the driving force
and the steady state potential of a Brownian particle in a thermal heat bath.
The force, or the drift coefficient, can be decomposed as the sum of the 
\spart part and the \apart part satisfying the condition \eqref{decomp2}. 
The decomposition method
reveals some aspects of nonequilibrium steady states. 
In the overdamped dynamics, any steady state is infinitely degenerate in
the sense that it is shared by the family of force fields of the form
\eqref{force_od}. The most probable points do not coincide with
the stable fixed points. In the underdamped dynamics, the correspondence
between the force and the steady state potential is restrictive. The
Boltzmann-type distribution is realized only in either an equilibrium system
or a nonequilibrium system driven by a velocity-dependent force such as a
magnetic force. The shifted Boltzmann distribution uniquely determines 
the corresponding force field under the stability requirement. The shifted
Boltzmann distribution is a connection between the overdamped dynamics and
the underdamped dynamics. 
As a byproduct, the decomposition method provides various 
examples of solvable nonequilibrium systems. 
Hopefully, they may be useful for the study of nonequilibrium statistical
mechanics.

\begin{acknowledgments}
This work was supported by the Basic Science Research Program through the
NRF Grant No.~2013R1A2A2A05006776.
\end{acknowledgments}

\appendix
%\bibliographystyle{apsrev}
%\bibliography{fd_noh}

\end{document}